\documentclass[twocolumn, amsfonts, amssymb, amsmath, pra]{revtex4-2}

\usepackage{latexsym}
\usepackage{amsthm}
\usepackage{graphicx}
\usepackage{ulem}
\usepackage{color}
\usepackage{natbib}
\usepackage{tensor}
\usepackage{mathrsfs}

\newcommand{\ket}[1]{ |{#1}\rangle}
\newcommand{\bra}[1]{ \langle{#1}|}
\newcommand{\tr}[0]{\mathrm{Tr}}
\newcommand{\hil}[1]{\mathbb{#1}}
\newcommand{\ESP}{\mathrm{ESP}}

\newcommand{\todai}{Department of Physics, Graduate School of Science, The University of Tokyo, Hongo 7-3-1, Bunkyo-ku, Tokyo 113-0033, Japan}

\newtheorem{lem}{Lemma}

\newtheorem{thm}{Theorem}

\allowdisplaybreaks[3]

\def\Proof{\textbf{Proof.} }

\begin{document}

\title{Optimal quantum discrimination of single-qubit unitary gates between two candidates}

\author{Akihito Soeda}
\email[Electronic address: ]{soeda@phys.s.u-tokyo.ac.jp}
\affiliation{\todai}

\author{Atsushi Shimbo}
\email[Electronic address: ]{shimbo@eve.phys.s.u-tokyo.ac.jp}
\affiliation{\todai}

\author{Mio Murao}
\email[Electronic address: ]{murao@phys.s.u-tokyo.ac.jp}
\affiliation{\todai}
\affiliation{Trans-scale Quantum Science Institute, The University of Tokyo, Hongo 7-3-1, Bunkyo-ku, Tokyo 113-0033, Japan}

\date{\today}

\begin{abstract}
We analyze a discrimination problem of a single-qubit unitary gate with two candidates, where the candidates are not provided with their classical description, but their quantum sample is.  More precisely, there are three unitary quantum gates--one target and one sample for each of the two candidates-- whose classical description is unknown except for their dimension.  The target gate is chosen equally among the candidates.  We obtain the optimal protocol that maximizes the expected success probability, assuming the Haar distribution for the candidates.
This problem is originally introduced in Ref.\,\cite{doi:10.1080/09500340903203129} which provides a protocol achieving 7/8 in the expected success probability based on the ``unitary comparison" protocol of Ref.\,\cite{Andersson_2003}.
The optimality of the protocol has been an open question since then.
We prove the optimality of the comparison protocol, implying that only one of the two samples (one for each candidate) is needed to achieve an optimal discrimination.
The optimization includes protocols outside the scope of quantum testers due to the dynamic ordering of the sample and target gates within a given protocol.
\end{abstract}

\maketitle

\section{Introduction}
A problem solving is an attempt to make the best judgement based on available resources.  In discrimination problems, the main goal is to correctly guess the identity of the target, provided that the target is chosen from a set of candidates with whatever information available regarding the candidates.

Quantum gates model time evolutions of quantum systems over a fixed duration.  The evolution may be a result of internal interaction of the system, or, perhaps, of a black-box device which transforms the state of the system via an evolution determined by the device.  There may be a physical operation performed at some distant location, say a ``quantum server" or an ``oracle", to which a quantum state is sent.  The state undergoes a prefixed evolution, and returns back to its origin.

Discrimination of quantum gates whose candidates are provided with their complete classical description reduces to finding the optimal initial state and measurement so that, when the target gate (\textit{i.e.}, the target of the discrimination) is applied, the possible output states, determined by which candidate the target resumes, become the most distinguishable.  (See Refs.\,\cite{Chefles2000,Bae_2015} and references therein for more on quantum state discrimination.)
For discrimination of unitary operations among two known candidates (\textit{i.e.}, with complete classical description), an explicit closed formula in terms of the overlap between the unitary operations is known for SU(2)~\cite{Ac_n_2001}.

\begin{figure}
    \includegraphics[width=8.5cm]{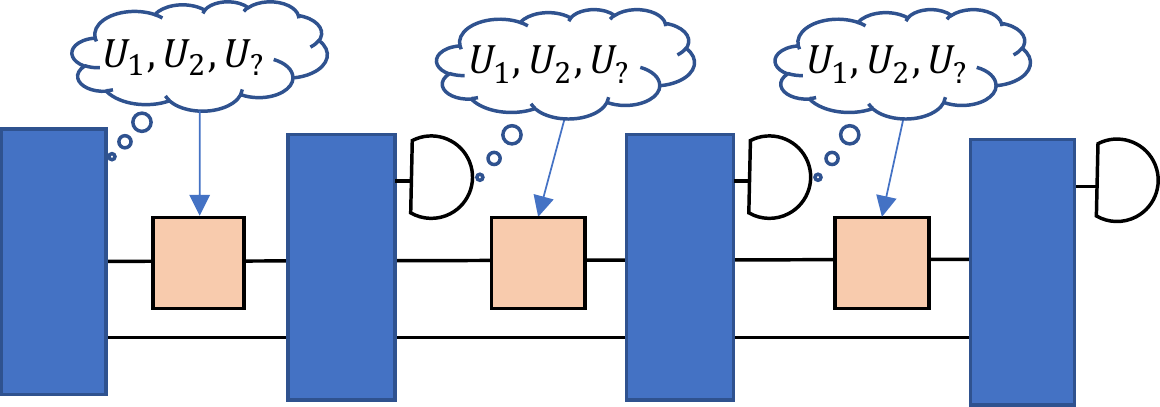}
    \caption{\label{fig1}Pictorial representation of our problem.  Pale orange boxes indicate the variable gates ($U_{[1]}$, $U_{[2]}$, and $U_{?}$), each implementing a single-qubit unitary operation whose action is not revealed to the discriminator.  Semicircle elements represent a quantum measurement, whose outcome can affect all the elements subsequent to it.  $U_{?}$ denotes the target of the discrimination, while $U_{[1]}$ and $U_{[2]}$ are a quantum sample of the first and second candidate, respectively.  More detailed descriptions are in the text and Fig.\,\ref{fig2}.}
\end{figure}

Quantum theory introduces a unique challenge to discrimination problems in that a full classical description of a quantum object cannot be deduced by physical operation in general when the number of copies of the quantum object is limited.
It is possible to introduce quantum process tomography to attempt to identify a complete classical description of the candidates and then proceed with the standard discrimination of known candidates.  (See Ref.\,\cite{Mohseni2008} and references therein for more on quantum process tomography.)
Quantum tomography may be unavoidable if the discrimination also requires us to identify a complete classical description of the target, but not so, if discrimination is more focused on particular properties of the target.

Suppose that a single-qubit unitary gate is given as a discrimination target, which is allowed to be used once.  The target is chosen between the two candidates with an equal probability.  A full classical description of the candidates is not available except that they are chosen independently and uniformly randomly from the set of all single-qubit unitary gates.  Instead of a classical description, we have one quantum gate as a ``quantum sample" for each candidate, implementing the unitary operation specified by the respective candidate.  Without any quantum sample of the candidates, discrimination would be impossible, while infinite access to the candidates would provide their complete classical descriptions.  We assume that the target and sample gates can be applied on any qubit and that a given gate has the same action on any qubit.  See Fig.\,\ref{fig1} for a pictorial description of our problem.

In this paper, we assume that each sample gate is allowed to be used once.  More precisely, there are three unitary quantum gates--one target and one sample for each of the two candidates-- whose classical description is unknown except for their dimension.  The performance of a given discrimination protocol will be measured by the expected probability of identifying the correct candidate.  We seek for the best performing protocol using a quantum system of an arbitrary size whose dynamics can be fully controlled, adopting a quantum circuit model as the basis of analysis.
We shall further specify the types of protocols to allow mathematically rigorous arguments but preserve the generality of the result.  The optimization will include all protocols of arbitrary circuit complexity.  Classical computation will be considered free.

This problem is originally introduced in Ref.\,\cite{doi:10.1080/09500340903203129} which provides a protocol that achieves 7/8 in the expected success probability based on the ``unitary comparison" protocol of Ref.\,\cite{Andersson_2003}.
This protocol ``compares" one of the quantum samples of the candidates against the target gate.  Thus the other quantum sample is discarded, but Ref.\,\cite{doi:10.1080/09500340903203129} reports to have not found any protocol outperforming the comparison.
The optimality of the protocol has been an open question since then.
Reference~\cite{Shimbo2018} derives 7/8 for discrimination protocols in which the sample and target gates are used in a predetermined order.
There still remained the possibility that more dynamic ordering of the gates depending on measurement outcomes used during a protocol can increase the expected success probability.
Recently, Ref.\,\cite{Wechs2021} numerically verifies that 7/8 is optimal, but analytic proof has been lacking.

The problem discussed above may be seen as a kind of ``programmable” discrimination of unitary gates, where the “reference” gates serve as a program.
The problem is also an instance of “pattern-matching” of unitary operations.
Analogous problem setting for quantum states are studied, for instance, in Refs.\,\cite{Barenco1997,Buhrman2001,Sasaki2001,Sasaki2002,Barnett2003,Jex2004,Carlini2005,Hayashi2005,Hayashi2006,Gu__2010,PhysRevA.94.062320,PhysRevLett.94.160501,HE2006103,PhysRevA.73.062334,PhysRevA.75.032316,PhysRevA.82.042312,Sentis2011,PhysRevA.89.014301,Sedlak2007,Ishida2008,Sedlak2008,Sedlak2009a,Olivares2011,Sentis2015,Jafarizadeh2017,Sentis2017,Liu2018,Sentis2019,Fanizza2019,Fanizza2020}.
Unitary comparison is further studied in Ref.\,\cite{Sedlak2009} with an unambiguousness condition.  Reference~\cite{Ziman2009} investigates unambiguous comparison of quantum measurements.
Identification of malfunctioning quantum devices \cite{Skotiniotis2018} and real-time calibration of optical receivers\cite{Bilkis2020} are more examples of more focused quantum process tomography.

Each quantum gate implements some quantum operation.
Discrimination of quantum operations with a \textit{complete} classical description of the candidates has been investigated for unitary operations \cite{Kitaev1997,Aharonov1998,Ac_n_2001,Acin2001,PhysRevLett.87.270404,PhysRevLett.98.100503,Zhou2007,PhysRevLett.101.180501,Duan2008,Zhang2008,Li2008,Wu2008,Laing2009,PhysRevLett.103.210501,Bisio2010a,Ziman2010,Hashimoto2010,Bisio2010,Reitzner2014,Bae2015,Cao2015a,Cao2016a,Li2017,Li2017a,Liu2019,Maffeis2019},
non-unitary deterministic quantum channels \cite{Kitaev1997,Sacchi_2005,PhysRevA.73.042301,PhysRevA.81.032339,PhysRevLett.101.180501,Bisio2009,Zhang2011,PhysRevLett.102.250501,Chiribella2012,Li2014,PhysRevLett.103.210501,Matthews2010,Ziman2008,Chiribella2012a,Puzzuoli2016,Rehman2018,Korzekwa2019,Wilde2020,Katariya2020,Bavaresco2020,Pereira2021},
quantum dynamics \cite{Childs2000,Aharonov2002,Wang2016,Yuan2017,Chen2019},
and stochastic quantum operations including quantum measurements \cite{Chefles2003,PhysRevLett.96.200401,Fiurasek2009,PhysRevA.90.052312,Mikova2014,Cao2015,Puchala2018}.
Discrimination of quantum dynamics is also studied in the context of quantum metrology (see Ref.\,\cite{Giovannetti2011} for review) which typically deals with continuously parametrized candidates.

Section \ref{prem} introduces notations used in this paper.  The problem setting and the main result are stated in Sec.\,\ref{setting}.  We argue that an upperbound to the optimal discrimination can be obtained by solving a semidefinite programming (SDP) problem in Sec.\,\ref{sec:primalSDP}.  A dual SDP problem is derived in Sec.\,\ref{sec:dualSDP}, which proves that 7/8 is an upperbound on the optimal performance, hence settling the optimality question of Ref.\,\cite{doi:10.1080/09500340903203129}.  We conclude in Sec.\,\ref{conclusion}.

\section{Preliminary} \label{prem}
Hilbert spaces are denoted by $\hil{H}$ with possible subscripts for distinction, \textit{e.g}, $\hil{H}_1$ and $\hil{H}_2$.  A vector will have a subscript to indicate the Hilbert space to which it belongs as in $\ket{\psi}_{\hil{H}}$.  Often, we will use the subscript of the Hilbert space to specify the corresponding Hilbert space as in $\ket{\psi}_1 \in \hil{H}_1$.  If a vector is in a tensor product of two or more Hilbert spaces, then the vector will have multiple subscripts as in $\ket{\varphi}_{12} \in \hil{H}_1 \otimes \hil{H}_2$.
Given a Hilbert space $\hil{H}$, we denote its \textit{copy}, \textit{i.e.}, another Hilbert space of the same dimension, with an overline as $\overline{\hil{H}}$.  The index of $\ket{\lambda}_{\overline{\hil{H}}_1}$ will be abbreviated as $\ket{\lambda}_{\overline{1}}$.
Similar conventions on subscripts will be adopted for operators and maps, throughout.
The Roman alphabet $I$ always represents the identity operator on its respective Hilbert space.
To each Hilbert space we designate a computational basis which will be identified with a tilde symbol above vectors as in $\ket{\widetilde{i}}$.
The symbol $\phi^+$ shall be reserved for the ``unnormalized" maximally entangled state as in
\begin{equation}
  \phi^+_{ab} = \sum_{i,j=0}^{d-1} \ket{\widetilde{i}}\bra{\widetilde{j}}_a \otimes \ket{\widetilde{i}}\bra{\widetilde{j}}_b,
\end{equation}
assuming $\hil{H}_a$ and $\hil{H}_b$ are both dimension $d$.

A deterministic quantum operation on a quantum system, identified by its corresponding Hilbert space $\hil{H}$, is described by a completely positive and trace-preserving (CPTP) map from $\mathcal{L}(\hil{H})$--the set of linear operators on $\hil{H}$-- to itself.  We denote the identity CPTP map from $\mathcal{L}(\hil{H})$ to itself as $\mathcal{I}_{\hil{H}}$.

A quantum channel takes the state of a given system to that of another system.  The state of the first system may have undergone a transformation, resulting in a different state in the latter system.
Thus, deterministic quantum operations and quantum channels share the same mathematical structure.  This motivates us to regard a quantum operation on a single quantum system also as a quantum channel.

A quantum channel may be implemented without a quantum measurement, but with measurements and addition and removal of subsystems, the set of possible quantum operations extends to \textit{quantum instruments}.  A quantum instrument $\mathscr{J}$ is characterized by an indexed set of completely positive maps $\mathscr{J} = \{ \mathcal{J}^{(i)} | \mathcal{J}^{(i)}: CP \}_i$ such that their sum $\mathcal{J} = \sum_i \mathcal{J}^{(i)}$ is trace-preserving.  The index $i$ is returned as a measurement outcome.

Given a unitary matrix $U$, $\mathcal{C}_{a \rightarrow b}( * ; U)$ is a unitary quantum channel from $\mathcal{L}(\hil{H}_a)$ to $\mathcal{L}(\hil{H}_b)$ defined by its action on the computational basis,
\[
\mathcal{C}_{a \rightarrow b}(\ket{\widetilde{k}}\bra{\widetilde{l}}_a ; U) = \sum_{k'l'} U[k',k] U[l,l']^* \ket{\widetilde{k'}}\bra{\widetilde{l'}}_b,
\]
 where $U[k',k]$ denotes the $(k',k)$ element of the unitary matrix $U$ and $U[l,l']^*$ the complex conjugate of the $(l,l')$ element.
In the case where $a = b$, we will use $\mathcal{C}_{a;U}$ as an abbreviation of $\mathcal{C}_{a \rightarrow b}( * ; U)$.

\section{Problem Setting and Main Result} \label{setting}
The sample gates and target gate constitute the ``variables" of the task.
Each variable gate may be applied on any single-qubit system.
The variable gates may be interpreted as a unitary channel $\mathcal{C}_{a \rightarrow b}(* ; U)$ for some suitable $U \in \text{SU}(2)$, the set of $2 \times 2$ special unitary matrices.
The corresponding $U$ depends on the computational basis of the qubit on which a variable gate applies, but we assume that the bases are so chosen that, given a particular variable gate, the corresponding unitary channel is always represented by the same unitary matrix.
This is, for instance, guaranteed by assuring that the ``up" direction of each spin is properly aligned, assuming spin-$\frac{1}{2}$'s are used for the qubits.
Let $U_{[1]}$ and $U_{[2]}$ correspond to the first and second candidate, respectively, while the unitary matrix of the target gate is either $U_{[1]}$ and $U_{[2]}$, depending on to which candidate the target corresponds.

Given a protocol $\mathcal{P}$, its probability of success $p_{\mathrm{suc}} (U_{[1]}, U_{[2]}, t, \mathcal{P})$ is determined by the action of the variable gates and the protocol $\mathcal{P}$, where $t = 1$ if the first candidate is chosen for the target, and $t = 2$, if the second.
The expected success probability $\ESP(\mathcal{P})$ of a given $\mathcal{P}$ is
\begin{equation}
  \ESP(\mathcal{P}) = \frac{1}{2} \sum_{t=1,2} \int p_{\mathrm{suc}} (U_{[1]}, U_{[2]}, t, \mathcal{P}) dU_{[1]} dU_{[2]},
\end{equation}
 where the integral for $U_{[1]}$ and $U_{[2]}$ is taken over the Haar measure of $\text{SU}(2)$.
The protocol must be independent of $U_{[1]}$, $U_{[2]}$, and $t$, as these are hidden at the beginning of the protocol.

We assume that a discrimination protocol always uses a finite-dimensional system and consists of a ``gate sequence".  A gate sequence in the quantum circuit model consists of applying quantum gates, which may implement any quantum operation, possibly nondeterministic, whose outcome is distinguished by a natural number.  The gates may also include any addition and removal of subsystems.  The outcomes of nondeterministic operations may affect the choice of subsequent operations, determined by classical computation on the preceding outcomes.  A protocol shall execute in a series of ``steps", where one gate is applied at each step, all subject to the outcomes of previous steps.  It terminates with producing a ``guess label", interpreted as its guess for the candidate gate to which the target gate corresponds.

If $\mathcal{P}$ outputs a label other than 1 and 2, then there always exists another protocol which assigns this label to either 1 or 2, only to increases from $\ESP(\mathcal{P})$.  The guess label is a function of all measurement outcomes produced during the protocol.  Ignoring circuit complexity, any protocol can always be redesigned without sacrificing $\ESP$ so that it terminates with a quantum measurement whose outcome is the guess label.  Without loss of generality, we optimize over protocols whose final step involves only a two-outcome measurement that outputs labels 1 and 2 and that this outcome is the guess label.

\begin{figure}
    \includegraphics[width=8.5cm]{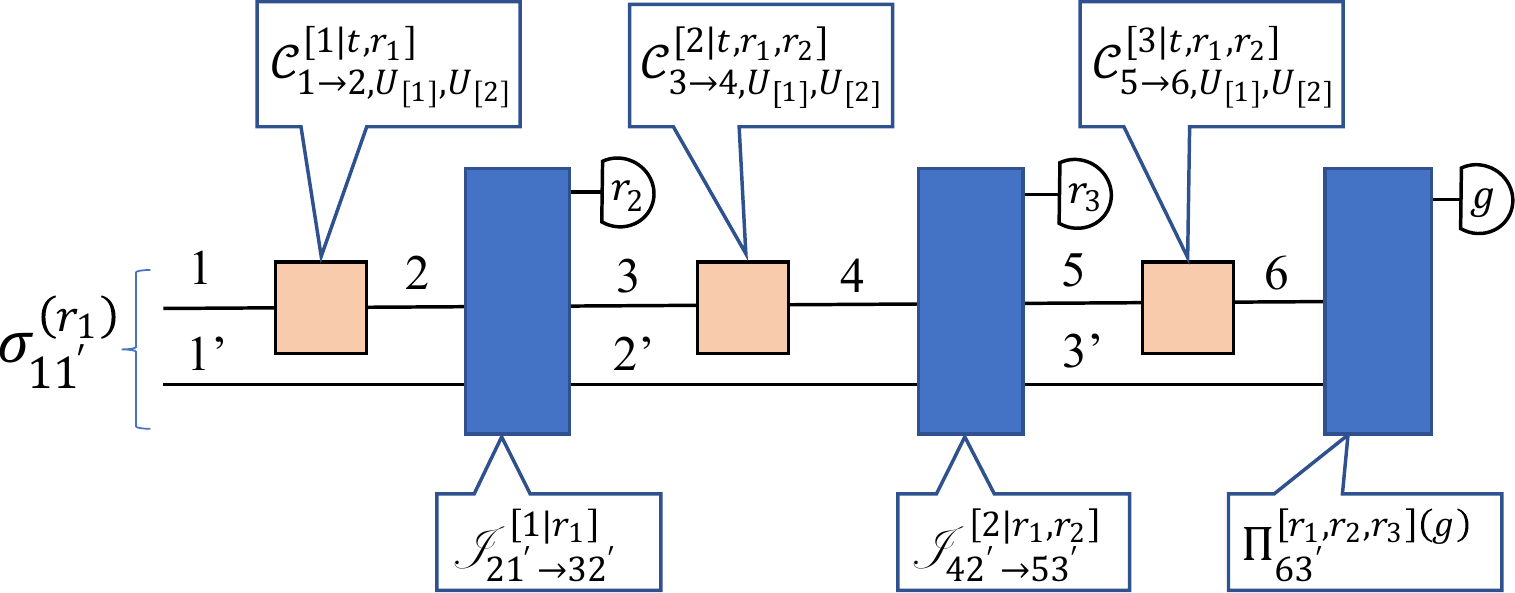}
    \caption{\label{fig2}Pictorial representation of the protocols $\mathcal{P}$ considered in this paper.  Pale orange boxes indicate the variable gates.  The state $\sigma^{(r_1)}_{11'}$ is chosen with probability $p^{(r_1)}$.  See Sec.\,\ref{setting} for exact description of each element.}
\end{figure}

We further assume that $\mathcal{P}$ proceeds in seven steps.  A pictorial representation is given in Fig.\,\ref{fig2}.
\begin{enumerate}
  \item Probabilistically prepare a quantum state labeled by $r_1$, according to a probability distribution $p^{(r_1)}$.  The quantum system used in this step must contain at least one qubit and the rest of the system be a finite-dimensional system.  The Hilbert space of this qubit is $\hil{H}_1$ and the rest of the system is $\hil{H}_{1'}$.  The state so prepared is ${\sigma'}^{(r_1)}_{11'}$.  It is more convenient that we introduced its unnormalized form including $p^{(r_1)}$, namely,
  \begin{equation}
    \sigma^{(r_1)}_{11'} := p^{(r_1)} {\sigma'}^{(r_1)}_{11'}.
  \end{equation}
  \item Choose one variable gate, depending on $r_1$, and apply it to the qubit corresponding to $\hil{H}_1$.
  We denote the unitary matrix corresponding to the chosen gate by $U_{U_{[1]}, U_{[2]}}(1|t,r_1)$.
  Set the input Hilbert space of this gate as $\hil{H}_1$ and the output as $\hil{H}_2$.  The action of the gate is given by
  \begin{equation}
    \mathcal{C}^{[1|t,r_1]}_{1 \rightarrow 2;U_{[1]}, U_{[2]}} (*) := \mathcal{C}_{1 \rightarrow 2}(*; U_{U_{[1]}, U_{[2]}}(1|t,r_1)).
  \end{equation}
  \item Apply a nondeterministic quantum operation, determined by $r_1$.  This operation returns an outcome $r_2$ and must leave at least one qubit in the system.  The Hilbert space of this qubit after the quantum operation is $\hil{H}_3$.  The Hilbert space of the rest of the system after the quantum system is $\hil{H}_{2'}$.
  This quantum operation is a quantum instrument from $\hil{H}_2 \otimes \hil{H}_{1'}$ to $\hil{H}_3 \otimes \hil{H}_{2'}$,
  \[
  \mathscr{J}_{21'\rightarrow 32'}^{[1|r_1]} = \{ \mathcal{J}_{21'\rightarrow 32'}^{[1|r_1](r_2)} \}_{r_2}.
  \]
  \item Choose one of the two remaining variable gates, depending on $r_1$ and $r_2$, and apply it to the qubit corresponding to $\hil{H}_3$.
  We denote the unitary matrix corresponding to the chosen gate by $U_{U_{[1]}, U_{[2]}}(2|t,r_1,r_2)$.
  Set the input Hilbert space of this gate as $\hil{H}_3$ and the output as $\hil{H}_4$.  The action of the gate is given by
  \begin{equation}
    \mathcal{C}^{[2|t,r_1,r_2]}_{3 \rightarrow 4;U_{[1]}, U_{[2]}} (*) := \mathcal{C}_{3 \rightarrow 4}(*; U_{U_{[1]}, U_{[2]}}(2|t,r_1,r_2)).
  \end{equation}
  \item Apply a nondeterministic quantum operation, determined by $r_1$ and $r_2$.  This operation returns an outcome $r_3$ and must leave at least one qubit in the system.  The Hilbert space of this qubit after the quantum operation is $\hil{H}_5$.  The Hilbert space of the rest of the system after the quantum system is $\hil{H}_{3'}$.
  This quantum operation is a quantum instrument from $\hil{H}_4 \otimes \hil{H}_{2'}$ to $\hil{H}_5 \otimes \hil{H}_{3'}$,
  \[
  \mathscr{J}_{42' \rightarrow 53'}^{[2|r_1,r_2]} = \{ \mathcal{J}_{42' \rightarrow 53'}^{[2|r_1,r_2](r_3)} \}_{r_3}.
  \]
  \item Apply the last variable gate to the qubit corresponding to $\hil{H}_5$.
  We denote the unitary matrix corresponding to the chosen gate by $U_{U_{[1]}, U_{[2]}}(3|t,r_1,r_2)$.  This is the last remaining variable gate hence independent of $r_3$.
  Set the input Hilbert space of this gate as $\hil{H}_5$ and the output as $\hil{H}_6$.  The action of the gate is given by
  \begin{equation}
    \mathcal{C}^{[3|t,r_1,r_2]}_{5 \rightarrow 6;U_{[1]}, U_{[2]}} (*) := \mathcal{C}_{5 \rightarrow 6}(*; U_{U_{[1]}, U_{[2]}}{(3|t,r_1,r_2)}).
  \end{equation}
  \item Applies a quantum measurement given by a positive-operator valued measure (POVM) on $\hil{H}_6 \otimes \hil{H}_{3'}$ with $g$ as its outcome.
  The choice of this POVM may depend on $r_1$, $r_2$, and $r_3$.  The elements of the POVM are denoted as $ \Pi^{[r_1,r_2,r_3](g)}_{63'}$, where $g$ is $1$ or $2$.
  The outcome $g$ is used as the guess label.
\end{enumerate}
Strictly speaking, we are asserting the universality of such a protocol under the standard quantum gate model. Proving the universality would require an additional set of axioms, which if they were to be justified calls for yet another set of arguments.  The line of argument must be continued indefinitely, hence the universality is asserted, instead.  In fact, the universality of a certain model can only be confirmed, empirically, as in the case of Turing machines in defining computable functions.
The protocols defined above lie outside the scope of quantum testers introduced in~Ref.~\cite{PhysRevLett.101.180501}, with a definite causal structure (\textit{i.e.}, under the quantum comb formalism), since the variable gates are chosen according to the measurement outcomes obtained at intermediate steps.

We state our main result as a theorem.
\begin{thm} \label{mainthm}
  The optimal $\ESP$ is $\frac{7}{8}$.
\end{thm}

\section{Upperbound on ESP by SDP} \label{sec:primalSDP}
We shall see that an upperbound to the optimal $\ESP$ can be formulated within semidefinite programming (SDP).
For that, we present a construction of another protocol $\mathcal{P}'$ that achieves the same $\ESP$ as any given $\mathcal{P}$.  $\mathcal{P}'$ has an arguably simpler causal structure in that the order in which the variable gates are used is fixed by a random variable at the beginning of $\mathcal{P}'$ unlike $\mathcal{P}$, which in general dynamically decides on the variable gates to apply only after subsequent measurements are performed.
The general strategy is to begin by obtaining an operator description of the protocol using the state-map duality as in between CPTP maps and Choi operators.  We then use these operator descriptions to find the necessary states and operations.

First, we take the maps and states of $\mathcal{P}$ and define
$\rho^{(r_1,r_2,r_3,g)}_{1\overline{2}3\overline{4}5\overline{6}}$ (\textit{i.e.}, the ``Choi" operator) by
\begin{multline}
  \rho^{(r_1,r_2,r_3,g)}_{1\overline{2}3\overline{4}5\overline{6}}
  := \tr_{63'} [\Pi^{([r_1,r_2,r_3](g)}_{63'} ( \mathcal{J}_{42' \rightarrow 53'}^{[2|r_1,r_2](r_3)}\\
   \circ \mathcal{J}_{21'\rightarrow 32'}^{[1|r_1](r_2)}  ) ( \sigma^{(r_1)}_{11'} \otimes \phi^+_{2\overline{2}} \otimes \phi^+_{4\overline{4}} \otimes \phi^+_{6\overline{6}}) ].
\end{multline}
We then define
\begin{multline}\label{twirling}
  \tilde{\rho}^{(r_1,r_2,r_3,g)}_{1\overline{2}3\overline{4}5\overline{6}}
  := \\
  \int
   (\mathcal{C}_{1;V} \otimes \mathcal{C}_{\overline{2};W} \otimes \mathcal{C}_{3;V} \otimes \mathcal{C}_{\overline{4};W} \otimes \mathcal{C}_{5;V}   \otimes \mathcal{C}_{\overline{6};W})\\
   (\rho^{(r_1,r_2,r_3,g)}_{1\overline{2}3\overline{4}5\overline{6}}) dV dW.
\end{multline}
We start from $\tilde{\rho}^{(r_1,r_2,r_3,g)}_{1\overline{2}3\overline{4}5\overline{6}}$ and compute the following operators and numbers,
\begin{align}
  \mu^{(r_1,r_2,r_3)}_{1\overline{2}3\overline{4}5\overline{6}} &:= \sum_g \tilde{\rho}^{(r_1,r_2,r_3,g)}_{1\overline{2}3\overline{4}5\overline{6}}, \label{mu}\\
  \nu^{(r_1,r_2,r_3)}_{1\overline{2}3\overline{4}5} &:= \frac{1}{2} \tr_{\overline{6}} [
  \mu^{(r_1,r_2,r_3)}_{1\overline{2}3\overline{4}5\overline{6}}],\\
  \xi^{(r_1,r_2)}_{1\overline{2}3\overline{4}5} &:= \sum_{r_3} \nu^{(r_1,r_2,r_3)}_{1\overline{2}3\overline{4}5},\\
  \zeta^{(r_1,r_2)}_{13} &:=
  \frac{1}{4}\tr_{\overline{24}5} \xi^{(r_1,r_2)}_{1\overline{2}3\overline{4}5},\\
  p^{(r_1,r_2)} &:= \tr_{13} \zeta^{(r_1,r_2)}_{13}.
\end{align}
We further define several rank 1 operators.  A consecutive sequence of numbers like 1234 will be abbreviated as $1...4$.
$\psi'_{5\overline{5}}$, $\psi_M$, and $\psi_{M'}$ appear below can be any fixed pure-state density operator (\textit{cf.} there may be multiple instances of $\mathcal{P}'$ for a given $\mathcal{P}$, in general.).  The rank 1 operators of interest are
\begin{multline}
  {\eta}^{(r_1,r_2,r_3)}_{1...6\overline{1}...\overline{6}M'}
   := \sum_{g,g'} \sqrt{\tilde{\rho}^{(r_1,r_2,r_3,g)}_{1\overline{2}3\overline{4}5\overline{6}}}  (\bigotimes_{k=1}^6 \phi^+_{k\overline{k}}) \sqrt{\tilde{\rho}^{(r_1,r_2,r_3,g')}_{1\overline{2}3\overline{4}5\overline{6}}}\\
    \otimes \ket{g}\bra{g'}_{M'},
\end{multline}
\begin{multline}
  {\eta'}^{(r_1,r_2,r_3)}_{1...6\overline{1}...\overline{6}M'} := \sqrt{\nu^{(r_1,r_2,r_3)}_{1\overline{2}3\overline{4}5}}  (\bigotimes_{k=1}^5 \phi^+_{k\overline{k}}) \sqrt{\nu^{(r_1,r_2,r_3)}_{1\overline{2}3\overline{4}5}}\\
   \otimes \phi^+_{6\overline{6}} \otimes \psi_{M'},
\end{multline}
\begin{multline}
  \tau^{(r_1,r_2)}_{1...5\overline{1}...\overline{5}M} := \sum_{r_3,r'_3}
\sqrt{\nu^{(r_1,r_2,r_3)}_{1\overline{2}3\overline{4}5}}  (\bigotimes_{k=1}^5 \phi^+_{k\overline{k}}) \sqrt{\nu^{(r_1,r_2,r'_3)}_{1\overline{2}3\overline{4}5}} \\
\otimes \ket{r_3}\bra{r'_3}_M,
\end{multline}
and
\begin{multline}
  {\tau'}^{(r_1,r_2)}_{1...5\overline{1}...\overline{5}M} := \sqrt{\zeta^{(r_1,r_2)}_{13}}\phi^+_{13\overline{13}}\sqrt{\zeta^{(r_1,r_2)}_{13}}\\
   \otimes \phi^+_{2\overline{2}} \otimes \phi^+_{4\overline{4}} \otimes \psi'_{5\overline{5}} \otimes \psi_M. \label{tau}
\end{multline}

\begin{lem} \label{lem1}
  There exists a unitary operator ${U'}^{(r_1,r_2,r_3)}_{\overline{1}2\overline{3}4\overline{5}6M'}$ such that
  \begin{equation} \label{puri1}
    {\eta}^{(r_1,r_2,r_3)}_{1...6\overline{1}...\overline{6}M'} = {U'}^{(r_1,r_2,r_3)}_{\overline{1}2\overline{3}4\overline{5}6M'}{\eta'}^{(r_1,r_2,r_3)}_{1...6\overline{1}...\overline{6}M'}({U'}^{(r_1,r_2,r_3)}_{\overline{1}2\overline{3}4\overline{5}6M'})^\dag.
  \end{equation}
\end{lem}
\noindent\Proof
Observe that
${\eta}^{(r_1,r_2,r_3)}_{1...6\overline{1}...\overline{6}M'}$ is a purification of $\mu^{(r_1,r_2,r_3)}_{1\overline{2}3\overline{4}5\overline{6}}$, while ${\eta'}^{(r_1,r_2,r_3)}_{1...6\overline{1}...\overline{6}M'}$ is that of $\nu^{(r_1,r_2,r_3)}_{1\overline{2}3\overline{4}5} \otimes I_{\overline{6}}$.
We have $\sum_g \Pi^{[r_1,r_2,r_3](g)}_{63'} = I_{63'}$ and $\tr_6[\phi^+_{6\overline{6}}] = I_{\overline{6}}$, thus
\begin{equation}
   \mu^{(r_1,r_2,r_3)}_{1\overline{2}3\overline{4}5\overline{6}} = \nu^{(r_1,r_2,r_3)}_{1\overline{2}3\overline{4}5} \otimes I_{\overline{6}}.
 \end{equation}
Therefore, there exists a unitary operator satisfying Eq.\,\eqref{puri1}. \qed

We see that quantities defined above are related.
\begin{lem} \label{eqReducSt}
  \begin{equation}
    \tr_5 \xi^{(r_1,r_2)}_{1\overline{2}3\overline{4}5} = \zeta^{(r_1,r_2)}_{13} \otimes I_{\overline{2}} \otimes I_{\overline{4}}.
  \end{equation}
\end{lem}

\noindent\Proof
We have
\begin{equation}
  \tr_5 \xi^{(r_1,r_2)}_{1\overline{2}3\overline{4}5} = \frac{1}{2}\tr_{\overline{4}5} [\xi^{(r_1,r_2)}_{1\overline{2}3\overline{4}5}] \otimes I_{\overline{4}}
\end{equation}
from the trace-preserving condition of $\mathscr{J}_{42' \rightarrow 53'}^{[2|r_1,r_2]}$ and $\tr_4[\phi^+_{4\overline{4}}] = I_{\overline{4}}$.  The symmetry induced from Eq.\,\eqref{twirling} implies that
\begin{equation}
  \tr_{\overline{4}5} [\xi^{(r_1,r_2)}_{1\overline{2}3\overline{4}5}] = \frac{1}{2} \tr_{\overline{24}5} [\xi^{(r_1,r_2)}_{1\overline{2}3\overline{4}5}] \otimes I_{\overline{2}}.
\end{equation}
Substituting this to the previous equation proves the claim.
\qed

\begin{lem} \label{lem3}
  There exists a unitary operator $U^{(r_1,r_2)}_{245\overline{135}M}$ such that
  \begin{equation} \label{puri2}
    {\tau}^{(r_1,r_2)}_{1...5\overline{1}...\overline{5}M} = U^{(r_1,r_2)}_{245\overline{135}M}{\tau'}^{(r_1,r_2)}_{1...5\overline{1}...\overline{5}M}(U^{(r_1,r_2)}_{245\overline{135}M})^\dag.
  \end{equation}
\end{lem}
\noindent\Proof
${\tau}^{(r_1,r_2)}_{1...5\overline{1}...\overline{5}M}$ is a purification of $\tr_5 [\xi^{(r_1,r_2)}_{1\overline{2}3\overline{4}5}]$.
${\tau'}^{(r_1,r_2)}_{1...5\overline{1}...\overline{5}M}$ is that of $\zeta^{(r_1,r_2)}_{13} \otimes I_{\overline{2}} \otimes I_{\overline{4}}$.
Therefore, ${\tau}^{(r_1,r_2)}_{1...5\overline{1}...\overline{5}M}$ and ${\tau'}^{(r_1,r_2)}_{1...5\overline{1}...\overline{5}M}$ are related by a unitary operation $U^{(r_1,r_2)}_{245\overline{135}M}$ due to Lemma~\ref{eqReducSt}.\qed

\noindent
Finally, $p^{(r_1,r_2)}$ is a probability distribution over $(r_1,r_2)$, because
\begin{equation}
  \sum_{r_1,r_2} \tr_{13} [ \zeta^{(r_1,r_2)}_{13} ] = 1.
\end{equation}

\begin{figure}
    \includegraphics[width=8.5cm]{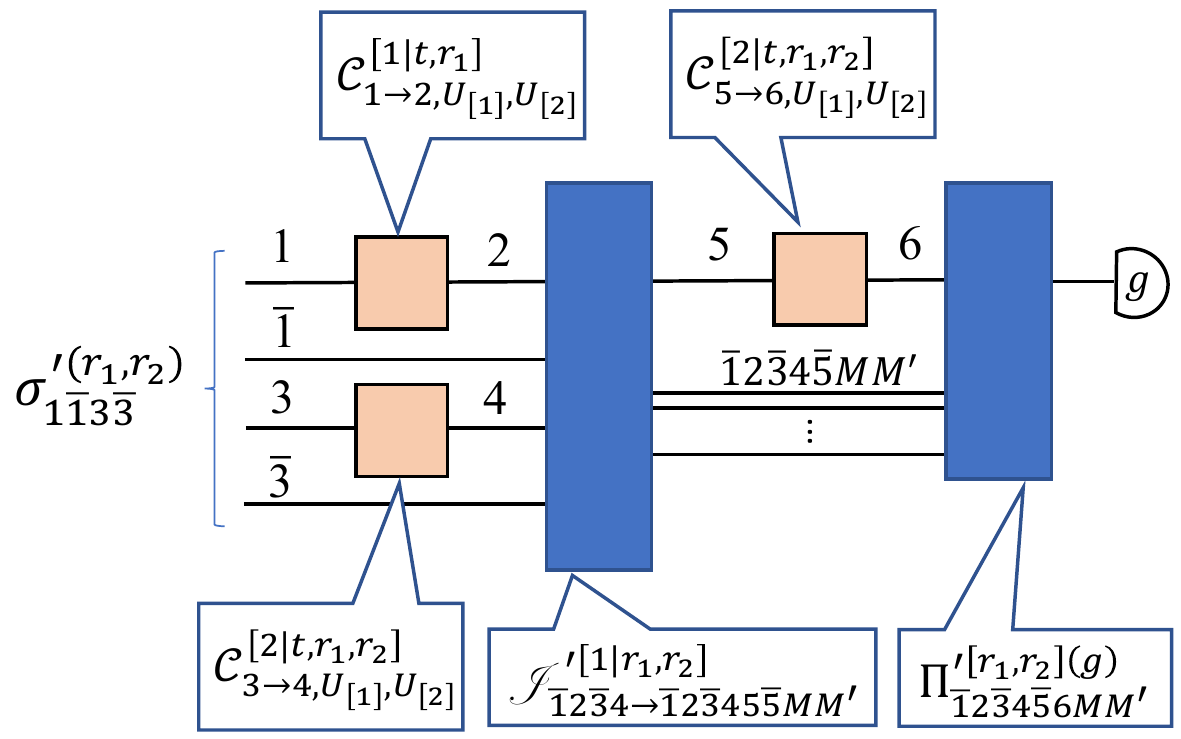}
    \caption{\label{fig3}Pictorial representation of $\mathcal{P}'$, including $\mathcal{P}^{\mathbf{M}}$.  All the elements of the protocol are chosen according to the random variables $r_1$ and $r_2$ chosen at the beginning of the protocol with probability $p^{(r_1,r_2)}$.  For $\mathcal{P}^{\mathbf{M}}$, $r_1$ and $r_2$ are fixed to $r^{\mathbf{M}}_1$ and $r^{\mathbf{M}}_2$, respectively.  See Sec.\,\ref{sec:primalSDP} for exact descriptions of each element.}
\end{figure}

We define the following protocol $\mathcal{P}'$ (Fig.~\ref{fig3}):
\begin{enumerate}
\item Prepare state
\begin{multline}
 {\sigma'}^{(r_1,r_2)}_{13\overline{13}} :=
  (p^{(r_1,r_2)})^{-1} \sqrt{\zeta^{(r_1,r_2)}_{13}}\phi^+_{13\overline{13}}\sqrt{\zeta^{(r_1,r_2)}_{13}}
\end{multline}
 with probability $p^{(r_1,r_2)}$.

 \item Choose $U_{U_{[1]}, U_{[2]}}(1|t,r_1)$ and $U_{U_{[1]}, U_{[2]}}(2|t,r_1,r_2)$ according to $r_1$ and $r_2$.  Apply the former to $\hil{H}_1$ and the latter to $\hil{H}_3$.

 \item Apply a quantum operation given by
 \begin{multline}
   {\mathcal{J}'}_{ \overline{1}2\overline{3}4 \rightarrow \overline{1}2\overline{3}45\overline{5}MM' }^{[r_1,r_2]} (*) :=\\
      U^{(r_1,r_2)}_{245\overline{135}M} (* \otimes \psi'_{5\overline{5}} \otimes \psi_M \otimes \psi_{M'}) (U^{(r_1,r_2)}_{245\overline{135}M})^\dag.
 \end{multline}

 \item Apply the final remaining variable gate $U_{U_{[1]}, U_{[2]}}(3|t,r_1,r_2)$ to $\hil{H}_5$.

 \item Perform a measurement given by $\{ {\Pi'}^{[r_1,r_2](g)}_{\overline{1}2\overline{3}4\overline{5}6MM'} \}_g$, where
 \begin{multline}
   {\Pi'}^{[r_1,r_2](g)}_{\overline{1}2\overline{3}4\overline{5}6MM'} \\
   = \sum_{r_3} ({U'}^{(r_1,r_2,r_3)}_{\overline{1}2\overline{3}4\overline{5}6M'})^\dag (I_{\overline{1}2\overline{3}4\overline{5}6} \otimes \ket{g}\bra{g}_{M'}){U'}^{(r_1,r_2,r_3)}_{\overline{1}2\overline{3}4\overline{5}6M'}\\
    \otimes \ket{r_3}\bra{r_3}_M.
 \end{multline}

 \item Declare $g$ as the guess of the candidate gate corresponding to the target gate.
\end{enumerate}
We define
\begin{multline}
  \tilde{\varrho}^{(r_1,r_2,g)}_{1\overline{2}3\overline{4}5\overline{6}}
    = \tr_{\overline{1}2\overline{3}4\overline{5}6MM'} [{\Pi'}^{[r_1,r_2](g)}_{\overline{1}2\overline{3}4\overline{5}6MM'}\\
    \qquad \cdot {\mathcal{J}'}_{ \overline{1}2\overline{3}4 \rightarrow \overline{1}2\overline{3}45\overline{5}MM' }^{[r_1,r_2]} ( {\sigma'}^{(r_1,r_2)}_{13\overline{13}} \otimes \phi^+_{2\overline{2}} \otimes \phi^+_{4\overline{4}} \otimes \phi^+_{6\overline{6}}) ].
\end{multline}

In $\mathcal{P}'$, the order of the variable gates is fixed once $(r_1, r_2)$ is chosen.  Nevertheless,
\begin{lem} \label{P=P'}
  $\mathcal{P}$ and $\mathcal{P}'$ achieve the same ESP.
\end{lem}

\noindent\Proof%
Introducing,
\begin{multline}
  {M'}^{[t,r_1,r_2]}_{1\overline{2}3\overline{4}5\overline{6};U_{[1]}, U_{[2]}} =
  ( \mathcal{C}^{[3|t,r_1,r_2]}_{5 \rightarrow 6;U_{[1]}, U_{[2]}} )^\dag
    \circ ( \mathcal{C}^{[2|t,r_1,r_2]}_{3 \rightarrow 4;U_{[1]}, U_{[2]}} )^\dag \\
    \qquad\qquad \circ (\mathcal{C}^{[1|t,r_1]}_{1 \rightarrow 2;U_{[1]}, U_{[2]}} )^\dag (\phi^+_{2\overline{2}} \otimes \phi^+_{4\overline{4}} \otimes \phi^+_{6\overline{6}}),
\end{multline}
we have for the original protocol
\begin{multline}
  p_{\mathrm{suc}} (U_{[1]}, U_{[2]}, t, \mathcal{P})\\
   = \sum_{r_1,r_2,r_3}  \tr_{1\overline{2}3\overline{4}5\overline{6}} [{M'}^{[t,r_1,r_2]}_{1\overline{2}3\overline{4}5\overline{6};U_{[1]}, U_{[2]}} {\rho}^{(r_1,r_2,r_3,t)}_{1\overline{2}3\overline{4}5\overline{6}}].
\end{multline}
The probability $p_{\mathrm{suc}} (U_{[1]}, U_{[2]}, t, \mathcal{P}')$ of correctly guessing the target gate with the new protocol is
\begin{multline}
  p_{\mathrm{suc}} (U_{[1]}, U_{[2]}, t, \mathcal{P}')\\
  =  \sum_{r_1,r_2} p^{(r_1,r_2)} \tr_{\overline{1}2\overline{3}4\overline{5}6MM'} [{\Pi'}^{[r_1,r_2](t)}_{\overline{1}2\overline{3}4\overline{5}6MM'} ( \mathcal{C}^{[3|t,r_1,r_2]}_{5 \rightarrow 6;U_{[1]}, U_{[2]}} \\
  \qquad\qquad \circ {\mathcal{J}'}_{ \overline{1}2\overline{3}4 \rightarrow \overline{1}2\overline{3}45\overline{5}MM' }^{[r_1,r_2]}
    \circ \mathcal{C}^{[2|t,r_1,r_2]}_{3 \rightarrow 4;U_{[1]}, U_{[2]}} \\
      \circ \mathcal{C}^{[1|t,r_1]}_{1 \rightarrow 2;U_{[1]}, U_{[2]}} ) ( {\sigma'}^{(r_1,r_2)}_{13\overline{13}}) ].
\end{multline}
Hence,
\begin{multline}
  p_{\mathrm{suc}} (U_{[1]}, U_{[2]}, t, \mathcal{P}')\\
   = \sum_{r_1,r_2} p^{(r_1,r_2)} \tr_{1\overline{2}3\overline{4}5\overline{6}} [{M'}^{[t,r_1,r_2]}_{1\overline{2}3\overline{4}5\overline{6};U_{[1]}, U_{[2]}} \tilde{\varrho}^{(r_1,r_2,t)}_{1\overline{2}3\overline{4}5\overline{6}}].
\end{multline}

 The states and operations of $\mathcal{P}'$ are chosen so that
\begin{align}
&p_{\mathrm{suc}} (U_{[1]}, U_{[2]}, t, \mathcal{P}')\\
&\qquad   = \sum_{r_1,r_2,r_3}  \tr_{1\overline{2}3\overline{4}5\overline{6}} [{M'}^{[t,r_1,r_2]}_{1\overline{2}3\overline{4}5\overline{6};U_{[1]}, U_{[2]}} \tilde{\rho}^{(r_1,r_2,r_3,t)}_{1\overline{2}3\overline{4}5\overline{6}}] \\
&\begin{aligned}\qquad   = \int \sum_{r_1,r_2,r_3,g}  \tr_{1\overline{2}3\overline{4}5\overline{6}} [{M'}^{[t,r_1,r_2]}_{1\overline{2}3\overline{4}5\overline{6};W^T U_{[1]}V, W^T U_{[2]} V} \\
\cdot \rho^{(r_1,r_2,r_3,g)}_{1\overline{2}3\overline{4}5\overline{6}}]  dV dW \end{aligned} \\
&\qquad   = \int p_{\mathrm{suc}} (W^T U_{[1]}V, W^T U_{[2]}V, t, \mathcal{P}) dV dW,
\end{align}
where $T$ denotes the transpose of a matrix.
To see the first equality, trace back the chain of definitions from Eq.\,\eqref{mu} to \eqref{tau} and also utilize Lemma \ref{lem1} and \ref{lem3}.
The ``off-diagonal" terms of $\eta$ and $\tau$ (\textit{i.e.}, for $r_3 \neq r'_3$ or $g \neq g'$) operators will disappear once the trace is taken over $M$ and then over $M'$.

Finally,
\begin{align}
&\ESP(\mathcal{P}') \\
&= \frac{1}{2} \sum_{t=1,2}  \int p_{\mathrm{suc}} (U_{[1]}, U_{[2]}, t, \mathcal{P}') dU_{[1]} dU_{[2]} \\
&\begin{aligned}
= \frac{1}{2} \sum_{t=1,2} \int \bigg( \int p_{\mathrm{suc}} (W^T U_{[1]}V, W^T U_{[2]}V, t, \mathcal{P}) \\ dU_{[1]} dU_{[2]} \bigg) dV dW \end{aligned}\\
&= \int \bigg(\frac{1}{2} \sum_{t=1,2} \int  p_{\mathrm{suc}} (U_{[1]}, U_{[2]}, t, \mathcal{P})  dU_{[1]} dU_{[2]} \bigg) dV dW \\
&= \int \ESP(\mathcal{P}) dV dW = \ESP(\mathcal{P}),
\end{align}
where the third to last equality follows from the group invariance of the Haar measure.
This proves that $\ESP(\mathcal{P})=\ESP(\mathcal{P}')$.
\qed

We are now ready to state the SDP problem that gives an upperbound to the optimal $\ESP$.
We define
\begin{align}
  \nonumber
  M^{\langle 1 \rangle}_{1\overline{2}3\overline{4}5\overline{6}} &:= \int
  (\mathcal{C}_{1;U} \otimes \mathcal{C}_{3;U} \otimes \mathcal{C}_{5;V})\\
  &\qquad\qquad\qquad(\phi^+_{1\overline{2}} \otimes \phi^+_{3\overline{4}} \otimes \phi^+_{5\overline{6}}) dU dV,\\
\nonumber
  M^{\langle 2 \rangle}_{1\overline{2}3\overline{4}5\overline{6}} &:= \int
  (\mathcal{C}_{1;U} \otimes \mathcal{C}_{3;V} \otimes \mathcal{C}_{5;U}) \\
  &\qquad\qquad\qquad(\phi^+_{1\overline{2}} \otimes \phi^+_{3\overline{4}} \otimes \phi^+_{5\overline{6}}) dU dV\\
\nonumber
  M^{\langle 3 \rangle}_{1\overline{2}3\overline{4}5\overline{6}} &:= \int
  (\mathcal{C}_{1;V} \otimes \mathcal{C}_{3;U} \otimes \mathcal{C}_{5;U}) \\
  &\qquad\qquad\qquad (\phi^+_{1\overline{2}} \otimes \phi^+_{3\overline{4}} \otimes \phi^+_{5\overline{6}}) dU dV,
\end{align}
which are the Choi operators of a certain random unitary channel as defined in each equation.
The SDP optimization of interest is given by
\begin{align}
\text{maximize}\quad & \frac{1}{2} \tr_{1\overline{2}3\overline{4}5\overline{6}} \left[ M^{\langle j \rangle}_{1\overline{2}3\overline{4}5\overline{6}} \rho^{(1)}_{1\overline{2}3\overline{4}5\overline{6}}  + M^{\langle j' \rangle}_{1\overline{2}3\overline{4}5\overline{6}} \rho^{(2)}_{1\overline{2}3\overline{4}5\overline{6}} \right] \label{p1} \\
\text{subject to}\quad & \rho^{(g)}_{1\overline{2}3\overline{4}5\overline{6}},  \rho^{\{1\}}_{1\overline{2}3\overline{4}5}, \rho^{\{2\}}_{13} \geq 0, \ g = 1,2 \label{p2}\\
& \sum_g \rho^{(g)}_{1\overline{2}3\overline{4}5\overline{6}} = \rho^{\{1\}}_{1\overline{2}3\overline{4}5} \otimes I_{\overline{6}} \label{p3} \\
& \tr_{5}[\rho^{\{1\}}_{1\overline{2}3\overline{4}5}]
= \rho^{\{2\}}_{13}  \otimes I_{\overline{2}} \otimes I_{\overline{4}} \label{p4} \\
& \tr_{13}[\rho^{\{2\}}_{13}] = 1 \label{p5}
\end{align}
for $\{j,j'\} = \{1,2\}, \{2,3\}, \{3,1\}$.

More formally stated:
\begin{lem} \label{ubbySDP}
  The SDP problem stated above gives an upperbound on the optimal ESP for at least one of the following assignments of $j$ and $j'$, namely, $\{j,j'\} = \{1,2\}, \{2,3\}, \{3,1\}$.
\end{lem}

To see this, first observe that the probability $p_{\mathrm{suc}} (U_{[1]}, U_{[2]}, t, \mathcal{P}'| r_1,r_2)$ of correctly guessing the target gate given $r_1$ and $r_2$ is
\begin{multline}
  p_{\mathrm{suc}} (U_{[1]}, U_{[2]}, t, \mathcal{P}'| r_1,r_2)\\
 =  \tr_{\overline{1}2\overline{3}4\overline{5}6MM'} [{\Pi'}^{[r_1,r_2](t)}_{\overline{1}2\overline{3}4\overline{5}6MM'} ( \mathcal{C}^{[3|t,r_1,r_2]}_{5 \rightarrow 6;U_{[1]}, U_{[2]}}\\
  \circ {\mathcal{J}'}_{ \overline{1}2\overline{3}4 \rightarrow \overline{1}2\overline{3}45\overline{5}MM' }^{(r_1,r_2)}
    \circ \mathcal{C}^{[2|t,r_1,r_2]}_{3 \rightarrow 4;U_{[1]}, U_{[2]}} \\
     \circ \mathcal{C}^{[1|t,r_1]}_{1 \rightarrow 2;U_{[1]}, U_{[2]}} ) ( {\sigma'}^{(r_1,r_2)}_{13\overline{13}}) ].
\end{multline}
Each $(r_1,r_2)$ corresponds a valid protocol.  Define
\begin{multline}
  \ESP(\mathcal{P}'| r_1,r_2) \\
  \qquad := \frac{1}{2} \sum_{t=1,2}  \int p_{\mathrm{suc}} (U_{[1]}, U_{[2]}, t, \mathcal{P}'| r_1,r_2) dU_{[1]} dU_{[2]}.
\end{multline}
Let $\mathcal{P}^{\mathbf{M}}$ be the protocol corresponding to $(r^{\mathbf{M}}_1,r^{\mathbf{M}}_2)$ that maximizes $\ESP(\mathcal{P}'| r_1,r_2)$.
Then,
\begin{equation}
  \ESP(\mathcal{P}')= \sum_{r_1,r_2} p^{(r_1,r_2)}\ESP(\mathcal{P}'| r_1,r_2)\leq \ESP(\mathcal{P}^{\mathbf{M}}).
\end{equation}

Let ${\sigma}^{\mathbf{M}}_{13\overline{13}}$, $\mathcal{J}_{ \overline{1}2\overline{3}4 \rightarrow \overline{1}2\overline{3}45\overline{5}MM' }^{\mathbf{M}}$, and $\{ {\Pi^{\mathbf{M}}}^{(g)}_{\overline{1}2\overline{3}4\overline{5}6MM'} \}$ be the corresponding operations used in $\mathcal{P}^{\mathbf{M}}$.
Define
$\rho^{\mathbf{M}(g)}_{1\overline{2}3\overline{4}5\overline{6}}$ by
\begin{multline}
  \rho^{\mathbf{M}(g)}_{1\overline{2}3\overline{4}5\overline{6}}
  := \tr_{\overline{1}2\overline{3}4\overline{5}6MM'} [{\Pi^{\mathbf{M}}}^{(g)}_{\overline{1}2\overline{3}4\overline{5}6MM'} \\
  \cdot \mathcal{J}_{ \overline{1}2\overline{3}4 \rightarrow \overline{1}2\overline{3}45\overline{5}MM' }^{\mathbf{M}} ( {\sigma}^{\mathbf{M}}_{13\overline{13}} \otimes \phi^+_{2\overline{2}} \otimes \phi^+_{4\overline{4}}) \otimes \phi^+_{6\overline{6}} ].
\end{multline}
With this,
\begin{equation} \label{ESPrho*}
  \ESP(\mathcal{P}^{\mathbf{M}}) = \frac{1}{2} \sum_{t=1,2} \tr_{1\overline{2}3\overline{4}5\overline{6}}
  [M^{[t|r^{\mathbf{M}}_1,r^{\mathbf{M}}_2]}_{1\overline{2}3\overline{4}5\overline{6}} \rho^{*(t)}_{1\overline{2}3\overline{4}5\overline{6}}],
\end{equation}
where
\begin{multline}
  M^{[t|r^{\mathbf{M}}_1,r^{\mathbf{M}}_2]}_{1\overline{2}3\overline{4}5\overline{6}} = \int
  ( \mathcal{C}^{[3|t,r^{\mathbf{M}}_1,r^{\mathbf{M}}_2]}_{5 \rightarrow 6;U_{[1]}, U_{[2]}} )^\dag
    \circ ( \mathcal{C}^{[2|t,r^{\mathbf{M}}_1,r^{\mathbf{M}}_2]}_{3 \rightarrow 4;U_{[1]}, U_{[2]}} )^\dag \\
    \qquad\qquad \circ (\mathcal{C}^{[1|t,r^{\mathbf{M}}_1]}_{1 \rightarrow 2;U_{[1]}, U_{[2]}} )^\dag (\phi^+_{2\overline{2}} \otimes \phi^+_{4\overline{4}} \otimes \phi^+_{6\overline{6}})
  dU_{[1]} dU_{[2]}
\end{multline}
with $\dag$ symbols indicating the adjoint map.
Regardless of the order of the variable gates and the value of $t$, $M^{[t|r^{\mathbf{M}}_1,r^{\mathbf{M}}_2]}_{1\overline{2}3\overline{4}5\overline{6}}$ is equal to one of $M^{\langle 1 \rangle}_{1\overline{2}3\overline{4}5\overline{6}}$, $M^{\langle 2 \rangle}_{1\overline{2}3\overline{4}5\overline{6}}$, or $M^{\langle 3 \rangle}_{1\overline{2}3\overline{4}5\overline{6}}$.
The operators $\rho^{\mathbf{M}(g)}_{1\overline{2}3\overline{4}5\overline{6}}$ satisfy the following relations
\begin{equation} \label{rho*1}
  \rho^{\mathbf{M}(g)}_{1\overline{2}3\overline{4}5\overline{6}} \geq 0
\end{equation}
and
\begin{equation} \label{rho*2}
 \sum_g \rho^{\mathbf{M}(g)}_{1\overline{2}3\overline{4}5\overline{6}} = \rho^{\{1\}}_{1\overline{2}3\overline{4}5} \otimes I_{\overline{6}},
\end{equation}
where $\rho^{\{1\}}_{1\overline{2}3\overline{4}5} = \frac{1}{2} \tr_{\overline{6}} [
\sum_g \rho^{\mathbf{M}(g)}_{1\overline{2}3\overline{4}5\overline{6}}]$.
In addition,
\begin{equation} \label{rho*3}
  \tr_{5}[\rho^{\{1\}}_{1\overline{2}3\overline{4}5}]
= \rho^{\{2\}}_{13}  \otimes I_{\overline{2}} \otimes I_{\overline{4}},
\end{equation}
where $\rho^{\{2\}}_{13} = \frac{1}{4}\tr_{\overline{24}5}
[\rho^{\{1\}}_{1\overline{2}3\overline{4}5}]$.  We also have that
\begin{equation} \label{rho*4}
  \tr_{13}[\rho^{\{2\}}_{13}] = 1.
\end{equation}
Notice that $\rho^{\{1\}}_{1\overline{2}3\overline{4}5}$ and $\rho^{\{2\}}_{13}$ are both positive operators.
This completes the proof of Lemma~\ref{ubbySDP}.  In principle, any optimal solution to the (primal) SDP problem gives a valid optimal discrimination protocol.

\section{Upper bound on the primal SDP} \label{sec:dualSDP}
\begin{lem} \label{lem:dualSDP}
An upperbound to the SDP \eqref{p1}-\eqref{p5} is given by the following SDP.
\begin{align}
\text{minimize}\quad & \lambda  \notag \\
\text{subject to}\quad & \Omega_{1\overline{2}3\overline{4}5\overline{6}},\Omega'_{1\overline{2}3\overline{4}}, \lambda \geq 0 \label{d1} \\
& \frac{1}{2} M^{\langle j \rangle}_{1\overline{2}3\overline{4}5\overline{6}} - \Omega_{1\overline{2}3\overline{4}5\overline{6}} \leq 0 \label{d2}\\
& \frac{1}{2} M^{\langle j' \rangle}_{1\overline{2}3\overline{4}5\overline{6}} - \Omega_{1\overline{2}3\overline{4}5\overline{6}} \leq 0 \label{d3}\\
& \tr_{\overline{6}}[ \Omega_{1\overline{2}3\overline{4}5\overline{6}} ] =  \Omega'_{1\overline{2}3\overline{4}} \otimes I_{5} \label{d4}\\
& \tr_{\overline{24}}[ \Omega'_{1\overline{2}3\overline{4}} ] = \lambda I_{1} \otimes I_3 \label{d5}
\end{align}
\end{lem}

To see this, the maximum on quantity \eqref{p1} is unaffected by adding constraints \eqref{d1}-\eqref{d5} because \eqref{p1} does not contain $\Omega$'s and so do constraints \eqref{p2} to \eqref{p5}.  One choice is to set $\Omega$'s proportional to the identity operators and set the proportionality constants large enough to satisfy conditions \eqref{d2} and \eqref{d3}.

Under the constraints \eqref{p2} to \eqref{p5} and \eqref{d1} to \eqref{d5}, quantity \eqref{p1} is equal to
\begin{align}
    L &:= \frac{1}{2} \tr_{1\overline{2}3\overline{4}5\overline{6}} \left[ M^{\langle j \rangle}_{1\overline{2}3\overline{4}5\overline{6}} \rho^{(1)}_{1\overline{2}3\overline{4}5\overline{6}}  + M^{\langle j' \rangle}_{1\overline{2}3\overline{4}5\overline{6}} \rho^{(2)}_{1\overline{2}3\overline{4}5\overline{6}} \right] \notag \\
    &\quad - \tr_{1\overline{2}3\overline{4}5\overline{6}} \left[
    \Omega_{1\overline{2}3\overline{4}5\overline{6}}
    (
    (\sum_g \rho^{(g)}_{1\overline{2}3\overline{4}5\overline{6}}) - \rho^{\{1\}}_{1\overline{2}3\overline{4}5} \otimes I_{\overline{6}}
     )
     \right] \notag \\
    &\quad - \tr_{1\overline{2}3\overline{4}} \left[ \Omega'_{1\overline{2}3\overline{4}}
    (
    \tr_{5}[\rho^{\{1\}}_{1\overline{2}3\overline{4}5}]
    - \rho^{\{2\}}_{13}  \otimes I_{\overline{2}} \otimes I_{\overline{4}}
    )\right]\notag \\
    &\quad - \lambda (\tr_{13}[\rho^{\{2\}}_{13}] - 1),
\end{align}

Rewriting $L$, we have
    \begin{align}
    L &= \tr_{1\overline{2}3\overline{4}5\overline{6}} \left[ \rho^{(1)}_{1\overline{2}3\overline{4}5\overline{6}} \left(
\frac{1}{2} M^{\langle j \rangle}_{1\overline{2}3\overline{4}5\overline{6}} - \Omega_{1\overline{2}3\overline{4}5\overline{6}}
    \right)
    \right] \notag\\
    &\quad + \tr_{1\overline{2}3\overline{4}5\overline{6}} \left[\rho^{(2)}_{1\overline{2}3\overline{4}5\overline{6}}\left(
    \frac{1}{2} M^{\langle j' \rangle}_{1\overline{2}3\overline{4}5\overline{6}} - \Omega_{1\overline{2}3\overline{4}5\overline{6}}
    \right)\right] \notag \\
    &\quad + \tr_{1\overline{2}3\overline{4}5} \left[ \rho^{\{1\}}_{1\overline{2}3\overline{4}5}       (\tr_{\overline{6}}[ \Omega_{1\overline{2}3\overline{4}5\overline{6}} ] -  \Omega'_{1\overline{2}3\overline{4}} \otimes I_{5}) \right] \notag \\
    &\quad + \tr_{13} \left[ \rho^{\{2\}}_{13} (\tr_{\overline{24}}[ \Omega'_{1\overline{2}3\overline{4}} ] - \lambda I_{1} \otimes I_3 )\right] \notag \\
    &\quad + \lambda.
    \end{align}
The constraints imply that $L \leq \lambda$.  Clearly, any such $\lambda$ is feasible if we remove constraints \eqref{p2}-\eqref{p5}.  Hence any feasible $\lambda$ under constraints \eqref{d1} to \eqref{d5} is an upper bound on all feasible values of quantity \eqref{p1} under constraints \eqref{p2}-\eqref{p5}.

The minimum attainable $\lambda$ depends on $j$ and $j'$ via constraints \eqref{d2} and \eqref{d3}.  At this point there are three possible combinations of $\{ j,j'\}$, namely, $\{ 1, 2 \}$, $\{ 2, 3 \}$, and $\{ 3, 1 \}$.  Combinations $\{ 1, 2 \}$ and $\{ 3, 1 \}$ reach the same minimum because they are related by swapping 1 with 3 and $\overline{2}$ with $\overline{4}$.

We provide explicit choices of $\lambda$, $\Omega$'s, and $\Omega'$'s that satisfy the constraints \eqref{d1}-\eqref{d5}, all with $\lambda = \frac{7}{8}$.  This shows that $\frac{7}{8}$ is an upperbound on the SDP in Lemma \ref{lem:dualSDP}.  The choices are expressed using the irreducible representation of $\text{SU}(2)$, which is motivated by the fact that $M^{\langle i \rangle}_{1\overline{2}3\overline{4}5\overline{6}}$'s are all block-diagonal in this basis as we see below.

We define the following basis of a three-qubit system $\hil{H}_{a} \otimes \hil{H}_{b} \otimes \hil{H}_{c} $,
\begin{align}
\ket{v_{\frac{1}{2}00}}_{abc} &= \frac{1}{\sqrt{2}} (\ket{\tilde{0}\tilde{1}\tilde{0}}_{abc} - \ket{\tilde{1}\tilde{0}\tilde{0}}_{abc}), \\
\ket{v_{\frac{1}{2}10}}_{abc} &= \frac{1}{\sqrt{2}} (\ket{\tilde{0}\tilde{1}\tilde{1}}_{abc} - \ket{\tilde{1}\tilde{0}\tilde{1}}_{abc}), \\
\ket{v_{\frac{1}{2}01}}_{abc} &= \sqrt{\frac{2}{3}} \ket{\tilde{0}\tilde{0}\tilde{1}}_{abc} - \sqrt{\frac{1}{6}} (\ket{\tilde{0}\tilde{1}\tilde{0}}_{abc} + \ket{\tilde{1}\tilde{0}\tilde{0}}_{abc}), \\
\ket{v_{\frac{1}{2}11}}_{abc} &= -\sqrt{\frac{2}{3}} \ket{\tilde{1}\tilde{1}\tilde{0}}_{abc} + \sqrt{\frac{1}{6}} (\ket{\tilde{0}\tilde{1}\tilde{1}}_{abc} + \ket{\tilde{1}\tilde{0}\tilde{1}}_{abc}),\\
\ket{v_{\frac{3}{2}0}}_{abc} &= \ket{\tilde{0}\tilde{0}\tilde{0}}_{abc}, \\
\ket{v_{\frac{3}{2}1}}_{abc} &= \sqrt{\frac{1}{3}}(\ket{\tilde{0}\tilde{1}\tilde{0}}_{abc} + \ket{\tilde{0}\tilde{0}\tilde{1}}_{abc} +\ket{\tilde{1}\tilde{0}\tilde{0}}_{abc}), \\
\ket{v_{\frac{3}{2}2}}_{abc} &= \sqrt{\frac{1}{3}}(\ket{\tilde{0}\tilde{1}\tilde{1}}_{abc} + \ket{\tilde{1}\tilde{1}\tilde{0}}_{abc} +\ket{\tilde{1}\tilde{0}\tilde{1}}_{abc}), \\
\ket{v_{\frac{3}{2}3}}_{abc} &= \ket{\tilde{1}\tilde{1}\tilde{1}}_{abc}.
\end{align}

We parametrize $M^{\langle 1 \rangle}_{1\overline{2}3\overline{4}5\overline{6}}$ as an example,
\begin{multline}
  M^{\langle 1 \rangle}_{1\overline{2}3\overline{4}5\overline{6}} = \\
   \big(\sum_{\substack{k_1,k_2, l_1, l'_1,\\ l_2, l'_2 = 0,1}} m^{(\frac{1}{2}\frac{1}{2})}_{{l_1} {l'_1} {l_2} {l'_2}} \ket{v_{\frac{1}{2}{k_1}{l_1}}}\bra{v_{\frac{1}{2}{k_1}{l'_1}}}_{135}\\
    \otimes \ket{v_{\frac{1}{2}{k_2}{l_2}}}\bra{v_{\frac{1}{2}{k_2}{l'_2}}}_{\overline{246}} \big) \\
   + \big(\sum_{k_1 = 0}^3 \sum_{k_2 = 0,1} \sum_{ l, l' = 0,1} m^{(\frac{3}{2}\frac{1}{2})}_{ll'} \ket{v_{\frac{3}{2}{k_1}}}\bra{v_{\frac{3}{2}{k_1}}}_{135} \\
   \otimes \ket{v_{\frac{1}{2}{k_2}{l}}}\bra{v_{\frac{1}{2}{k_2}{l'}}}_{\overline{246}} \big)\\
   + \big(\sum_{k_1 = 0,1} \sum_{k_2 = 0}^3 \sum_{ l, l' = 0,1} m^{(\frac{1}{2}\frac{3}{2})}_{ll'} \ket{v_{\frac{1}{2}{k_1}{l}}}\bra{v_{\frac{1}{2}{k_1}{l'}}}_{135} \\
   \otimes \ket{v_{\frac{3}{2}{k_2}}}\bra{v_{\frac{3}{2}{k_2}}}_{\overline{246}} \big)\\
   + \big(\sum_{k_1 = 0}^3 \sum_{k_2 = 0}^3  m^{(\frac{3}{2}\frac{3}{2})} \ket{v_{\frac{3}{2}{k_1}}}\bra{v_{\frac{1}{2}{k_1}}}_{135} \\
    \otimes \ket{v_{\frac{3}{2}{k_2}}}\bra{v_{\frac{3}{2}{k_2}}}_{\overline{246}} \big).
\end{multline}
In this parametrization,
\begin{align}
& \left(
\begin{array}{cccc}
  m^{(\frac{1}{2}\frac{1}{2})}_{0000} & m^{(\frac{1}{2}\frac{1}{2})}_{0001} & m^{(\frac{1}{2}\frac{1}{2})}_{0010} & m^{(\frac{1}{2}\frac{1}{2})}_{0011} \\
  m^{(\frac{1}{2}\frac{1}{2})}_{0100} & m^{(\frac{1}{2}\frac{1}{2})}_{0101} & m^{(\frac{1}{2}\frac{1}{2})}_{0110} & m^{(\frac{1}{2}\frac{1}{2})}_{0111} \\
  m^{(\frac{1}{2}\frac{1}{2})}_{1000} & m^{(\frac{1}{2}\frac{1}{2})}_{1001} & m^{(\frac{1}{2}\frac{1}{2})}_{1010} & m^{(\frac{1}{2}\frac{1}{2})}_{1011} \\
  m^{(\frac{1}{2}\frac{1}{2})}_{1100} & m^{(\frac{1}{2}\frac{1}{2})}_{1101} & m^{(\frac{1}{2}\frac{1}{2})}_{1110} & m^{(\frac{1}{2}\frac{1}{2})}_{1111} \\
\end{array}
\right) = \left(
\begin{array}{cccc}
 \frac{1}{2} & 0 & 0 & 0 \\
 0 & 0 & 0 & 0 \\
 0 & 0 & 0 & 0 \\
 0 & 0 & 0 & \frac{1}{6} \\
\end{array}
\right), \\
&\qquad\qquad\left(
\begin{array}{cc}
 m^{(\frac{3}{2}\frac{1}{2})}_{00} & m^{(\frac{3}{2}\frac{1}{2})}_{01} \\
 m^{(\frac{3}{2}\frac{1}{2})}_{10} & m^{(\frac{3}{2}\frac{1}{2})}_{11} \\
\end{array}
\right) = \left(
\begin{array}{cc}
 0 & 0 \\
 0 & \frac{1}{6} \\
\end{array}
\right),\\
&\qquad\qquad\left(
\begin{array}{cc}
 m^{(\frac{1}{2}\frac{3}{2})}_{00} & m^{(\frac{1}{2}\frac{3}{2})}_{01} \\
m^{(\frac{1}{2}\frac{3}{2})}_{10} & m^{(\frac{1}{2}\frac{3}{2})}_{11} \\
\end{array}
\right) = \left(
\begin{array}{cc}
 0 & 0 \\
 0 & \frac{1}{6} \\
\end{array}
\right),\\
&\qquad\qquad\qquad\qquad m^{(\frac{3}{2}\frac{3}{2})} = \frac{1}{6}.
\end{align}
The other $M^{\langle i \rangle}_{1\overline{2}3\overline{4}5\overline{6}}$'s are obtained by suitably swapping the qubits.

Next, we introduce the basis of a two-qubit system $\hil{H}_{a} \otimes \hil{H}_{b}$ as
\begin{align}
\ket{w_{0}}_{ab} &= \frac{1}{\sqrt{2}}(\ket{\tilde{0}\tilde{1}}_{ab} - \ket{\tilde{1}\tilde{0}}_{ab}),\\
\ket{w_{10}}_{ab} &= \ket{\tilde{0}\tilde{0}}_{ab},\\
\ket{w_{11}}_{ab} &= \frac{1}{\sqrt{2}}(\ket{\tilde{0}\tilde{1}}_{ab} + \ket{\tilde{1}\tilde{0}}_{ab}),\\
\ket{w_{12}}_{ab} &= \ket{\tilde{1}\tilde{1}}_{ab}.
\end{align}
We parametrize $\Omega_{1\overline{2}3\overline{4}5\overline{6}}$ and $\Omega'_{1\overline{2}3\overline{4}}$ as
\begin{multline}
  \Omega_{1\overline{2}3\overline{4}5\overline{6}} = \\
   \big(\sum_{\substack{k_1,k_2, l_1, l'_1,\\ l_2, l'_2 = 0,1}}
    \omega^{(\frac{1}{2}\frac{1}{2})}_{{l_1} {l'_1} {l_2} {l'_2}} \ket{v_{\frac{1}{2}{k_1}{l_1}}}\bra{v_{\frac{1}{2}{k_1}{l'_1}}}_{135} \\
     \otimes \ket{v_{\frac{1}{2}{k_2}{l_2}}}\bra{v_{\frac{1}{2}{k_2}{l'_2}}}_{\overline{246}} \big) \\
   + \big(\sum_{k_1 = 0}^3 \sum_{k_2 = 0,1} \sum_{ l, l' = 0,1} \omega^{(\frac{3}{2}\frac{1}{2})}_{ll'} \ket{v_{\frac{3}{2}{k_1}}}\bra{v_{\frac{3}{2}{k_1}}}_{135} \\
   \otimes \ket{v_{\frac{1}{2}{k_2}{l}}}\bra{v_{\frac{1}{2}{k_2}{l'}}}_{\overline{246}} \big)\\
   + \big(\sum_{k_1 = 0,1} \sum_{k_2 = 0}^3 \sum_{ l, l' = 0,1} \omega^{(\frac{1}{2}\frac{3}{2})}_{ll'} \ket{v_{\frac{1}{2}{k_1}{l}}}\bra{v_{\frac{1}{2}{k_1}{l'}}}_{135} \\
   \otimes \ket{v_{\frac{3}{2}{k_2}}}\bra{v_{\frac{3}{2}{k_2}}}_{\overline{246}} \big)\\
   + \big(\sum_{k_1 = 0}^3 \sum_{k_2 = 0}^3  \omega^{(\frac{3}{2}\frac{3}{2})} \ket{v_{\frac{3}{2}{k_1}}}\bra{v_{\frac{1}{2}{k_1}}}_{135} \\
    \otimes \ket{v_{\frac{3}{2}{k_2}}}\bra{v_{\frac{3}{2}{k_2}}}_{\overline{246}} \big)
\end{multline}
\begin{multline}
  \Omega'_{1\overline{2}3\overline{4}} = \\
   ( {\omega'}^{(00)} \ket{w_0}\bra{w_0}_{135} \otimes \ket{w_0}\bra{w_0}_{\overline{246}} )\\
   + (\sum_{k_2 = 0}^2  {\omega'}^{(01)} \ket{w_0}\bra{w_0}_{135} \otimes \ket{w_{1{k_2}}}\bra{w_{1{k_2}}}_{\overline{246}} )\\
   + (\sum_{k_1 = 0}^2  {\omega'}^{(10)} \ket{w_{1{k_1}}}\bra{w_{1{k_1}}}_{135} \otimes \ket{w_0}\bra{w_0}_{\overline{246}} )\\
   + (\sum_{k_1=0}^2 \sum_{k_2=0}^2 {\omega'}^{(11)} \ket{w_{1{k_1}}}\bra{w_{1{k_1}}}_{135} \otimes \ket{w_{1{k_2}}}\bra{w_{1{k_2}}}_{\overline{246}} ).
\end{multline}
For $j=1$ and $j'=2$, a feasible set of parameters of the dual SDP is
\begin{align}
&\qquad\qquad\qquad\lambda = \frac{7}{8}, \\
&\left(
\begin{array}{cccc}
  \omega^{(\frac{1}{2}\frac{1}{2})}_{0000} & \omega^{(\frac{1}{2}\frac{1}{2})}_{0001} & \omega^{(\frac{1}{2}\frac{1}{2})}_{0010} & \omega^{(\frac{1}{2}\frac{1}{2})}_{0011} \\
  \omega^{(\frac{1}{2}\frac{1}{2})}_{0100} & \omega^{(\frac{1}{2}\frac{1}{2})}_{0101} & \omega^{(\frac{1}{2}\frac{1}{2})}_{0110} & \omega^{(\frac{1}{2}\frac{1}{2})}_{0111} \\
  \omega^{(\frac{1}{2}\frac{1}{2})}_{1000} & \omega^{(\frac{1}{2}\frac{1}{2})}_{1001} & \omega^{(\frac{1}{2}\frac{1}{2})}_{1010} & \omega^{(\frac{1}{2}\frac{1}{2})}_{1011} \\
  \omega^{(\frac{1}{2}\frac{1}{2})}_{1100} & \omega^{(\frac{1}{2}\frac{1}{2})}_{1101} & \omega^{(\frac{1}{2}\frac{1}{2})}_{1110} & \omega^{(\frac{1}{2}\frac{1}{2})}_{1111} \\
\end{array}
\right) =
\left(
\begin{array}{cccc}
 \frac{1}{4} & 0 & 0 & 0 \\
 0 & \frac{1}{16} & \frac{1}{16} & \frac{1}{8 \sqrt{3}} \\
 0 & \frac{1}{16} & \frac{1}{16} & \frac{1}{8 \sqrt{3}} \\
 0 & \frac{1}{8 \sqrt{3}} & \frac{1}{8 \sqrt{3}} & \frac{1}{6} \\
\end{array}
\right), \\
&\qquad\left(
\begin{array}{cc}
 \omega^{(\frac{3}{2}\frac{1}{2})}_{00} & \omega^{(\frac{3}{2}\frac{1}{2})}_{01} \\
 \omega^{(\frac{3}{2}\frac{1}{2})}_{10} & \omega^{(\frac{3}{2}\frac{1}{2})}_{11} \\
\end{array}
\right) = \left(
\begin{array}{cc}
 \frac{1}{16} & -\frac{1}{16 \sqrt{3}} \\
 -\frac{1}{16 \sqrt{3}} & \frac{5}{48} \\
\end{array}
\right),\\
&\qquad\left(
\begin{array}{cc}
 \omega^{(\frac{1}{2}\frac{3}{2})}_{00} & \omega^{(\frac{1}{2}\frac{3}{2})}_{01} \\
\omega^{(\frac{1}{2}\frac{3}{2})}_{10} & \omega^{(\frac{1}{2}\frac{3}{2})}_{11} \\
\end{array}
\right) = \left(
\begin{array}{cc}
 \frac{1}{16} & -\frac{1}{16 \sqrt{3}} \\
 -\frac{1}{16 \sqrt{3}} & \frac{1}{6} \\
\end{array}
\right),\\
&\qquad\qquad\omega^{(\frac{3}{2}\frac{3}{2})} = \frac{5}{48},~{\omega'}^{(00)} = \frac{1}{2},~ {\omega'}^{(01)} = \frac{1}{8},\\
&\qquad\qquad~{\omega'}^{(10)} = \frac{1}{8},~ {\omega'}^{(11)} = \frac{1}{4}.
\end{align}
A feasible set of parameters of the dual SDP for the other pair ($j=2$ and $j'=3$) is
\begin{align}
&\qquad\qquad\qquad\lambda = \frac{7}{8}, \\
& \left(
\begin{array}{cccc}
  \omega^{(\frac{1}{2}\frac{1}{2})}_{0000} & \omega^{(\frac{1}{2}\frac{1}{2})}_{0001} & \omega^{(\frac{1}{2}\frac{1}{2})}_{0010} & \omega^{(\frac{1}{2}\frac{1}{2})}_{0011} \\
  \omega^{(\frac{1}{2}\frac{1}{2})}_{0100} & \omega^{(\frac{1}{2}\frac{1}{2})}_{0101} & \omega^{(\frac{1}{2}\frac{1}{2})}_{0110} & \omega^{(\frac{1}{2}\frac{1}{2})}_{0111} \\
  \omega^{(\frac{1}{2}\frac{1}{2})}_{1000} & \omega^{(\frac{1}{2}\frac{1}{2})}_{1001} & \omega^{(\frac{1}{2}\frac{1}{2})}_{1010} & \omega^{(\frac{1}{2}\frac{1}{2})}_{1011} \\
  \omega^{(\frac{1}{2}\frac{1}{2})}_{1100} & \omega^{(\frac{1}{2}\frac{1}{2})}_{1101} & \omega^{(\frac{1}{2}\frac{1}{2})}_{1110} & \omega^{(\frac{1}{2}\frac{1}{2})}_{1111} \\
\end{array}
\right) = \left(
\begin{array}{cccc}
 \frac{1}{16} & 0 & 0 & \frac{1}{16} \\
 0 & \frac{1}{8} & \frac{1}{8} & 0 \\
 0 & \frac{1}{8} & \frac{1}{8} & 0 \\
 \frac{1}{16} & 0 & 0 & \frac{11}{48} \\
\end{array}
\right), \\
&\qquad\qquad\left(
\begin{array}{cc}
 \omega^{(\frac{3}{2}\frac{1}{2})}_{00} & \omega^{(\frac{3}{2}\frac{1}{2})}_{01} \\
 \omega^{(\frac{3}{2}\frac{1}{2})}_{10} & \omega^{(\frac{3}{2}\frac{1}{2})}_{11} \\
\end{array}
\right) = \left(
\begin{array}{cc}
 \frac{1}{8} & 0 \\
 0 & \frac{1}{24} \\
\end{array}
\right),\\
&\qquad\qquad\left(
\begin{array}{cc}
 \omega^{(\frac{1}{2}\frac{3}{2})}_{00} & \omega^{(\frac{1}{2}\frac{3}{2})}_{01} \\
\omega^{(\frac{1}{2}\frac{3}{2})}_{10} & \omega^{(\frac{1}{2}\frac{3}{2})}_{11} \\
\end{array}
\right) = \left(
\begin{array}{cc}
 \frac{1}{8} & 0 \\
 0 & \frac{1}{24} \\
\end{array}
\right),\\
&\qquad\qquad\omega^{(\frac{3}{2}\frac{3}{2})} = \frac{13}{96},~{\omega'}^{(00)} = \frac{1}{8},~ {\omega'}^{(01)} = \frac{1}{4},\\
&\qquad\qquad{\omega'}^{(10)} = \frac{1}{4},~ {\omega'}^{(11)} = \frac{5}{24}.
\end{align}

\section{Conclusion} \label{conclusion}
We analyzed the discrimination of a single-qubit unitary gate with two candidates, whose complete classical descriptions are unknown but provided with one quantum sample for each candidate.  The target gate is chosen equally among the candidates.  The expected success probability (ESP) was chosen as the figure of merit.  We assumed the Haar distribution for the candidates.
This problem, originally introduced in Ref.\,\cite{doi:10.1080/09500340903203129}, is known to achieve at least 7/8 in ESP.
We proved that this indeed is the optimal by deriving an upperbound of the optimal ESP as a semidefinite programming (SDP) problem and providing explicit feasible parameters for its dual SDP.
Thus, we confirmed that the optimal discrimination is achievable without one of the quantum samples of the candidates.
The optimization covers all the protocols allowing dynamic ordering of the variable gates depending on measurement outcome obtained at the intermediate steps of a given protocol, thus going beyond the quantum testers.

\section*{Acknowledgment}
This work is supported by JSPS KAKENHI (Grant No.\,17H01694, No.\,18H04286, No.\,18K13467, and No.\,21H03394) and  MEXT Quantum Leap Flagship Program (MEXT Q-LEAP) Grant No.\,JPMXS0118069605).

\end{document}